\documentclass[aps,print,a4paper,showpacs,12pt,onecolumn]{revtex4}
\usepackage{amssymb}
\usepackage{amsmath}
\usepackage{graphicx}

\input{tcilatex}
\begin{document}

\title{Interaction of a nonlinear spin wave and magnetic soliton in a
uniaxial anisotropic ferromagnet}
\author{Zai-Dong Li$^{1}$, Qiu-Yan Li$^{1}$, Zhan-Guo Bai$^{2}$, and Yubao
Sun$^{1}$}
\affiliation{$^{1}$Department of Applied Physics, Hebei University of Technology, Tianjin
300130, China\\
$^{2}$College of Nature, Hebei University of Science and Technology,
Shijiazhuang 050054, China}

\begin{abstract}
We study the interaction of a nonlinear spin-wave and magnetic soliton in a
uniaxial anisotropic ferromagnet. By means of a reasonable assumption and a
straightforward Darboux transformation one- and two-soliton solutions in a
nonlinear spin-wave background are obtained analytically, and their
properties are discussed in detail. In the background of a nonlinear spin
wave the amplitude of the envelope soliton has the spatial and temporal
period, and soliton can be trapped only in space. The amplitude and wave
number of spin wave have the different contribution to the width, velocity,
and the amplitude of soliton solutions, respectively. The envelope of
solution hold the shape of soliton, and the amplitude of each envelope
soliton keeps invariability before and after collision which shows the
elastic collision of two envelope soliton in the background of a nonlinear
spin wave.
\end{abstract}

\maketitle

\section{Introduction}

Solitons are the natural results of the nonlinear equation in which the
dispersion is compensated by nonlinear effects. With this compensation
solitons can travel over long distances with neither attenuation nor change
of shape which has the prominent application in high-rate telecommunications
with optical fibers in the future. For this reason the study of solitons has
received considerable attentions in many fields, such as particle physics,
molecular biology, nonlinear optics \cite{Kivshar1} and condensate physics 
\cite{Trullinger,Kivshar2,Ablowitz}. One of the topics in condensate physics
is the nonlinear dynamics of local magnetization in Heisenberg spin chain
model which successfully explains the existence of ferromagnetism and
antiferromagnetism at temperatures below the Curie temperature. Taking into
account the spin-spin interactions, the nonlinear excitations, such as spin
wave and magnetic solitons \cite{Kosevich,Mikeska}, are general phenomena in
ordered ferromagnetic materials. By means of the neutron inelastic
scattering \cite{Boucher} and electron spin resonance \cite{Asano}, the
magnetic soliton had already been probed experimentally in
quasi-one-dimensional magnetic systems. The magnetic soliton, which
describes localized magnetization, is an important excitation in the
classical Heisenberg spin chain. In particular, the continuum limit for the
nonlinear dynamics of magnetization in the classical ferromagnet is governed
by the Landau-Lifshitz (L-L) equation \cite{landau}. This equation governs a
classical nonlinear dynamically system with novel properties. In a
one-dimensional case, some types of L-L equation is complete integrable. The
isotropic case has been studied in various aspects \cite%
{Laksmanan,Tjon,Takhtajan}, and the construction of soliton solutions of
anisotropic L-L equation is also discussed \cite%
{Bolovik,Fogedby,lizd1,Liu,Huang,liqy}. In these studies a variety of
techniques, such as inverse scattering transformation \cite%
{Takhtajan,Bolovik,Fogedby,lizd1}, Riemann-Hilbert approach \cite{Liu}, and
Darboux transformation \cite{Huang,liqy}, have been applied to construct the
soliton solution of L-L equation. Recently, this Heisenberg model has new
application in some fields of condensate physics. In magnetic multilayers
the magnetization switching and reversal \cite{Sun}, domain-wall dynamics 
\cite{S. Zhang} and magnetic solitons \cite{lizd2,pbhe} were discussed
analytically in terms of L-L equation with spin-torque. Taking into account
the light-induced and the magnetic dipole-dipole interactions the
ferromagnetic state \cite{Pu} are obtained and the nonlinear excitations,
such as spin wave \cite{Zhang} and magnetic solitons \cite{lizd3}, has been
reported in spinor Bose-Einstein condensates in deep optical lattice.

In this paper, we will study the interaction of a nonlinear spin wave and
magnetic solitons in a uniaxial anisotropic ferromagnet. In terms of a
reasonable assumption we transform the Landau-Lifshitz equation into an
equation of the nonlinear type. By means of a straightforward Darboux
transformation we report one- and two-soliton solutions in a nonlinear
spin-wave background analytically and discuss their properties in detail.

\section{Model}

In the classical limit, the dynamics of magnetization of a ferromagnet as a
function of space and time $\boldsymbol{M}\left( x,t\right) $ is determined
by the Landau-Lifshitz equation \cite{landau}%
\begin{equation}
\frac{\partial \mathbf{M}}{\partial t}=-\frac{2\mu _{0}}{\hbar }\mathbf{%
M\times H}_{eff},  \label{dynamics}
\end{equation}%
where $\mu _{0}$ is Bohr magneton, the effective magnetic field $\mathbf{H}%
_{eff}$ is equal to the variational derivative of the magnetic crystal
energy with respect to the vector $\mathbf{M}$, 
\begin{equation}
\mathbf{H}_{eff}=-\frac{\delta E}{\delta \mathbf{M}},  \label{field}
\end{equation}%
where the magnetic crystal energy function including the exchange energy,
anisotropic energy and the Zeeman energy can be written as \cite{Kosevich} 
\begin{equation}
E=\frac{1}{2}\int \left[ A\left( \frac{\partial \mathbf{M}}{\partial x}%
\right) ^{2}-\beta M_{z}^{2}-\mathbf{M\cdot B}\right] d^{3}x,  \label{energy}
\end{equation}%
where $A$ is the exchange constant and $\beta $ is the uniaxial anisotropic
constant, $\beta >0$ corresponds to easy-axis anisotropy while $\beta <0$
corresponds to easy-plane type. Substituting Eqs. (\ref{field}) and (\ref%
{energy}) into Eq. (\ref{dynamics}), we get%
\begin{equation}
\frac{\hbar }{2\mu _{0}}\frac{\partial \mathbf{M}}{\partial t}=-A\mathbf{%
M\times }\frac{\partial ^{2}\mathbf{M}}{\partial x^{2}}-\beta \mathbf{%
M\times e}_{3}(\mathbf{M\cdot e}_{3})-\mu _{0}\mathbf{M\times B},
\label{LL1}
\end{equation}%
where $\mathbf{e}_{3}$ is the unit vector along the $z$-axis, and $\mathbf{B}%
=(0,0,B)$. When the magnetic field is high enough, the deviation of
magnetization from the the direction of the field is small. With the
consideration that at temperatures well below the Curie temperature the
magnitude of magnetization is a constant, i.e., the integral of motion $%
\mathbf{M}^{2}\equiv M_{0}^{2}=constant$, here $M_{0}$ is the saturation
magnetization, we can introduce a single function $\Psi $, instead of two
independent components of $\mathbf{M}$,%
\begin{equation}
\Psi =m_{x}+im_{y},\text{ }m_{z}=\sqrt{1-\left\vert \Psi \right\vert ^{2}},
\label{note}
\end{equation}%
where $\mathbf{m\equiv }\left( m_{x},m_{y},m_{z}\right) =\mathbf{M}%
\boldsymbol{/}M_{0}$. Thus Eq. (\ref{LL1}) becomes%
\begin{equation}
\frac{i\hbar }{2\mu _{0}}\frac{\partial \Psi }{\partial t}-AM_{0}\left( m_{z}%
\frac{\partial ^{2}}{\partial x^{2}}\Psi -\Psi \frac{\partial ^{2}}{\partial
x^{2}}m_{z}\right) =-\left( \beta M_{0}m_{z}+B\right) \Psi .  \label{nls1}
\end{equation}

In the ground state, vector $\mathbf{M}$ directs along the anisotropy axis $%
\mathbf{e}_{3}$. Now we consider the small deviations of magnetization from
the equilibrium direction which corresponding to $m_{x}^{2}+m_{y}^{2}\ll
m_{z}^{2}$, i.e., $\left\vert \psi \right\vert ^{2}\ll 1$, then $%
m_{z}\approx 1-1/2\left\vert \Psi \right\vert ^{2}$. In the long-wavelength
approximation and the case $\beta >0$, Eq. (\ref{nls1}) may be simplified by
keeping only the nonlinear terms of the order of the magnitude of $%
\left\vert \psi \right\vert ^{2}\psi $. As a result, we have the following
dimensionless Schr\"{o}dinger equation:

\begin{equation}
i\frac{\partial \Psi }{\partial t}-\frac{1}{2}\frac{\partial ^{2}\Psi }{%
\partial x^{2}}-\left\vert \Psi \right\vert ^{2}\Psi +2(1+\frac{B}{\beta
M_{0}})\Psi =0.  \label{nls2}
\end{equation}%
For convenience we have rescaled space $x$ and time $t$ by $2l_{0}$ and $%
1/\omega _{0}$, where $l_{0}=\sqrt{A/\beta }$ is the characteristic magnetic
length and $\omega _{0}=\beta \mu _{0}M_{0}/\hbar $ is the homogeneous
ferromagnetic resonance frequency. The spin wave interaction in a
ferromagnet with easy-axis anisotropy is attractive in nature. This
attraction leads to the macroscopic phenomena which are associated with the
appearance of spatially localized magnetic excitations, i.e., magnetic
solitons which is admitted in nonlinear equation (\ref{nls2}). So it is very
interesting to investigate the dynamics of magnetic soliton in a nonlinear
spin wave background which hasn't been well explored. To this purpose we
have to get soliton solutions of (\ref{nls2}) embedded in a nonlinear
spin-wave background. It should be noted that Darboux transformation \cite%
{Matveev,Gu,Lilu} has been developed to construct such solutions. The main
idea of this method is that it firstly transforms the nonlinear equation
into the Lax representation, and then in terms of a series of reasonable
transformations the soliton solutions can be constructed algebraically with
an obvious seed solution of the nonlinear equation. By employing
Ablowitz-Kaup-Newell-Segur technique we can construct Lax representation for
Eq. (\ref{nls2}) as follows

\begin{eqnarray}
\frac{\partial }{\partial x}\psi &=&U\psi ,  \notag \\
\frac{\partial }{\partial t}\psi &=&V\psi ,  \label{laxE}
\end{eqnarray}%
where $\psi =\left( 
\begin{array}{cc}
\psi _{1} & \psi _{2}%
\end{array}%
\right) ^{T}$, the superscript \textquotedblleft $T$\textquotedblright\
denotes the matrix transpose. The lax pairs $U$ and $V$ are given in the
forms%
\begin{eqnarray}
U &=&\lambda J+P,  \notag \\
V &=&\left( -i\lambda ^{2}+\alpha _{2}\right) J-i\lambda P+\frac{1}{2}%
i(P^{2}+\frac{\partial }{\partial x}P)J,  \label{laxpair}
\end{eqnarray}%
with%
\begin{equation*}
J=\left( 
\begin{array}{ll}
1 & 0 \\ 
0 & -1%
\end{array}%
\right) ,\text{ }P=\left( 
\begin{array}{ll}
0 & \Psi \\ 
-\overline{\Psi } & 0%
\end{array}%
\right) ,\alpha _{2}=i(1+\frac{B}{\beta M_{0}}),
\end{equation*}%
where the overbar denotes the complex conjugate. With the natural condition
of Eq. (\ref{laxE}) $\frac{\partial ^{2}}{\partial x\partial t}\psi =\frac{%
\partial ^{2}}{\partial t\partial x}\psi $, i.e., $\frac{\partial }{\partial
t}U-\frac{\partial }{\partial x}V+\left[ U,V\right] =0$, the Eq. (\ref{nls2}%
) can be recovered. Based on the Eq. (\ref{laxE}), we can obtain the general
one- and two-soliton solution by using a straightforward Darboux
transformation \cite{Matveev,Gu,Lilu}.

\section{Darboux transformation}

In order to clear the correction of soliton solutions we briefly introduce
the procedure of getting soliton solutions for the developed Darboux
transformation. We consider the following transformation 
\begin{equation}
\psi \left[ 1\right] =\left( \lambda I-S\right) \psi ,  \label{Darboux1}
\end{equation}%
where $S=K\Lambda K^{-1},$ $\Lambda =$diag$\left( \lambda _{1},\lambda
_{2}\right) ,$ and $K$\hspace{0in} is a nonsingular matrix which satisfies 
\begin{equation}
K_{x}=JK\Lambda +PK.  \label{Darboux2}
\end{equation}%
Letting $\psi \left[ 1\right] $ satisfy the Lax equation 
\begin{equation}
\frac{\partial }{\partial x}\psi \left[ 1\right] =U_{1}\psi \left[ 1\right] ,
\label{Darboux3}
\end{equation}%
where $U_{1}=\lambda J+P_{1}$, $P_{1}=\left( 
\begin{array}{ll}
0 & \Psi _{1} \\ 
-\overline{\Psi }_{1} & \text{ \hspace{0in} \hspace{0in} }0%
\end{array}%
\right) $, and with the help of Eqs. (\ref{laxpair}), (\ref{Darboux1}) and (%
\ref{Darboux2}), we obtain the Darboux transformation for Eq. (\ref{nls2})
from Eq. (\ref{Darboux3}) in the form 
\begin{equation}
\Psi _{1}=\Psi +2S_{12}.  \label{Darboux4}
\end{equation}%
The above equation implies that a new solution of Eq. (\ref{nls2}) can be
obtained if $S$ is known. To obtain the expression of $S$ it is is easy to
verify that, if $\psi =\left( 
\begin{array}{cc}
\psi _{1} & \psi _{2}%
\end{array}%
\right) ^{T}$ is a eigenfunction of Eq. (\ref{laxE}) corresponding to the
eigenvalue $\lambda =\lambda _{1}$, then$\left( 
\begin{array}{cc}
-\overline{\psi }_{2} & \overline{\psi }_{1}%
\end{array}%
\right) ^{T}$ is also the eigenfunction, while with the eigenvalue $-%
\overline{\lambda }_{1}$. Therefore, $K$ and $\Lambda $ can be taken the
form 
\begin{equation}
K=\left( 
\begin{array}{ll}
\psi _{1} & -\overline{\psi }_{2} \\ 
\psi _{2} & \text{ }\overline{\psi }_{1}%
\end{array}%
\right) ,\Lambda =\left( 
\begin{array}{ll}
\lambda _{1} & \text{ }0 \\ 
0 & -\overline{\lambda }_{1}%
\end{array}%
\right) ,  \label{Darboux5}
\end{equation}%
which ensures that Eq. (\ref{Darboux2}) is held, we can obtain 
\begin{equation}
S_{sl}=-\overline{\lambda }_{1}\delta _{sl}+\left( \lambda _{1}+\overline{%
\lambda }_{1}\right) \frac{\psi _{s}\overline{\psi }_{l}}{\psi ^{T}\overline{%
\psi }},\text{ }s,l=1,2,  \label{Darboud6}
\end{equation}%
where $\psi ^{T}\overline{\psi }=\left\vert \psi _{1}\right\vert
^{2}+\left\vert \psi _{2}\right\vert ^{2}$. Then Eq. (\ref{Darboux4})
becomes 
\begin{equation}
\Psi _{1}=\Psi +2\left( \lambda _{1}+\overline{\lambda }_{1}\right) \frac{%
\psi _{1}\overline{\psi }_{2}}{\psi ^{T}\overline{\psi }},  \label{Darboux}
\end{equation}%
where $\psi =\left( 
\begin{array}{cc}
\psi _{1} & \psi _{2}%
\end{array}%
\right) ^{T}$ to be determined is the eigenfunction of Eq. (\ref{laxE})
corresponding to the eigenvalue $\lambda _{1}$ for the solution $\Psi $.
Thus if Eq. (\ref{laxE}) is solved we can generate a new solution $\Psi _{1}$%
, therefore the new solutions of $\mathbf{m}$ from Eq. (\ref{note}), of the
Eq. (\ref{nls2}) from a known solution $\Psi $ which is usually called
\textquotedblleft seed\textquotedblright\ solution.

To obtain exact $N$-order solution of Eq. (\ref{nls2}), we firstly rewrite
the Darboux transformation in Eq. (\ref{Darboux}) as in the form 
\begin{equation}
\Psi _{1}=\Psi +2\left( \lambda _{1}+\overline{\lambda }_{1}\right) \frac{%
\psi _{1}\left[ 1,\lambda _{1}\right] \overline{\psi }_{2}\left[ 1,\lambda
_{1}\right] }{\psi \left[ 1,\lambda _{1}\right] ^{T}\overline{\psi }\left[
1,\lambda _{1}\right] },  \label{Oneso}
\end{equation}%
where $\psi \left[ 1,\lambda \right] =\left( \psi _{1}\left[ 1,\lambda %
\right] ,\psi _{2}\left[ 1,\lambda \right] \right) ^{T}$ denotes the
eigenfunction of Eq. (\ref{laxpair}) corresponding to eigenvalue $\lambda $.
Then repeating above the procedure $N$ times, we can obtain the exact $N$%
-order solution 
\begin{equation}
\Psi _{N}=\Psi +2\sum_{n=1}^{N}(\lambda _{n}+\overline{\lambda }_{n})\frac{%
\psi _{1}[n,\lambda _{n}]\overline{\psi }_{2}[n,\lambda _{n}]}{\psi \lbrack
n,\lambda _{n}]^{T}\overline{\psi }[n,\lambda _{n}]},  \label{Multiso}
\end{equation}%
where 
\begin{align*}
\psi \left[ n,\lambda \right] & =\left( \lambda -S\left[ n-1\right] \right)
\cdots \left( \lambda -S\left[ 1\right] \right) \psi \left[ 1,\lambda \right]
, \\
S_{sl}\left[ j\right] & =-\overline{\lambda }_{j}\delta _{sl}+\left( \lambda
_{j}+\overline{\lambda }_{j}\right) \frac{\psi _{s}\left[ j,\lambda _{j}%
\right] \overline{\psi }_{l}\left[ j,\lambda _{j}\right] }{\psi \left[
j,\lambda _{j}\right] ^{T}\overline{\psi }\left[ j,\lambda _{j}\right] },
\end{align*}%
here $\psi \left[ j,\lambda \right] $ is the eigenfunction corresponding to $%
\lambda _{j}$ for $\Psi _{j-1}$ with $\Psi _{0}\equiv \Psi $ and $s,l=1,2$, $%
j=1,2,\cdots ,n-1$, $n=2,3,\cdots ,N$. Thus if choosing a \textquotedblleft
seed\textquotedblright\ as the basic initial solution, by solving linear
characteristic equation system (\ref{laxE}), one can construct a set of new
solutions of $\mathbf{m}$ from Eqs. (\ref{note}), (\ref{nls2}) and (\ref%
{Multiso}).

In order to get soliton solutions in a nonlinear spin wave background we
take the initial \textquotedblleft seed\textquotedblright\ solution as \ 
\begin{equation}
\Psi _{c}=A_{c}e^{-i(k_{c}x\mathbf{-}\omega _{c}t)},  \label{ansaz1}
\end{equation}%
which corresponding to a nonlinear spin wave solution of $\mathbf{m}$ as%
\begin{eqnarray}
m_{x} &=&A_{c}\cos \left( k_{c}x\mathbf{-}\omega _{c}t\right) ,  \notag \\
m_{y} &=&-A_{c}\sin \left( k_{c}x\mathbf{-}\omega _{c}t\right) ,  \notag \\
m_{z} &=&\sqrt{1-A_{c}^{2}},  \label{magnon}
\end{eqnarray}%
$\allowbreak $which satisfies the nonlinear dispersion relation $\omega
_{c}=k_{c}^{2}/2-A_{c}^{2}+2B/\left( \beta M_{0}\right) +2$, where the
amplitude $A_{c}$ is small real constants and $k_{c}$ is the wave number of
nonlinear spin wave. Substituting (\ref{ansaz1}) into (\ref{nls2}) and
solving the linear equation system (\ref{laxE}), after the tedious
calculation we have the eigenfunction of Eq. (\ref{laxE}) corresponding to
eigenvalue $\lambda $ in the form%
\begin{align}
\psi _{1}& =LC_{1}e^{\Theta _{1}}+A_{c}C_{2}e^{\Theta _{2}},  \notag \\
\psi _{2}& =A_{c}C_{1}e^{-\Theta _{2}}+LC_{2}e^{-\Theta _{1}},
\label{Laxsolution}
\end{align}%
$\allowbreak \allowbreak $where the parameters $C_{1}$ and $C_{2}$ are the
arbitrary complex constants, and the other parameters are defined by 
\begin{eqnarray*}
\Theta _{1} &=&-\frac{1}{2}i\left( k_{c}x-\omega _{c}t\right) +D\left(
x+\delta t\right) , \\
\Theta _{2} &=&-\frac{1}{2}i\left( k_{c}x-\omega _{c}t\right) -D\left(
x+\delta t\right) , \\
L &=&-\frac{1}{2}ik_{c}-D-\lambda , \\
D &=&\frac{1}{2}\sqrt{\left( ik_{c}+2\lambda \right) ^{2}-4A_{c}^{2}}, \\
\delta &=&-i\lambda -\frac{1}{2}k_{c},
\end{eqnarray*}%
Following the formulas (\ref{note}), (\ref{Multiso}), and (\ref{Laxsolution}%
) we can obtain the one- and two-soliton solutions embedded in a nonlinear
spin wave background, respectively.$\allowbreak $

\section{Modulation of one-soliton solution by a nonlinear spin wave}

Taking the spectral parameter $\lambda \equiv \lambda _{1}=\mu _{1}/2+i\nu
_{1}/2$, here $\mu _{1}$ and $\nu _{1}$ are real number, in Eq. (\ref%
{Laxsolution}) and substituting them into Eq. (\ref{Oneso}), we obtain the
one-soliton solution from Eqs. (\ref{note}) and (\ref{Multiso}) as follows%
\begin{eqnarray}
m_{x} &=&A_{c}\cos \varphi +\frac{\mu _{1}}{\Delta _{1}}\left( Q_{2}\sin
\varphi +Q_{1}\cos \varphi \right) ,  \notag \\
m_{y} &=&-A_{c}\sin \varphi +\frac{\mu _{1}}{\Delta _{1}}\left( Q_{2}\cos
\varphi -Q_{1}\sin \varphi \right) ,  \notag \\
m_{z} &=&\sqrt{1-\left( A_{c}+\frac{\mu _{1}}{\Delta _{1}}Q_{1}\right)
^{2}-\left( \frac{\mu _{1}}{\Delta _{1}}Q_{2}\right) ^{2}},
\label{onesoliton}
\end{eqnarray}%
where%
\begin{eqnarray}
\varphi &=&k_{c}x-\omega _{c}t,  \notag \\
\theta _{1} &=&2D_{1R}x+2\left( D_{1}\delta _{1}\right) _{R}t+2x_{0},  \notag
\\
\Phi _{1} &=&2D_{1I}x+2\left( D_{1}\delta _{1}\right) _{I}t-2\varphi _{0},
\label{para1}
\end{eqnarray}%
\begin{eqnarray}
D_{1} &=&\sqrt{\left( ik_{c}/2+\lambda _{1}\right) ^{2}-A_{c}^{2}},  \notag
\\
L_{1} &=&-ik_{c}/2-D_{1}-\lambda _{1},  \notag \\
\delta _{1} &=&-i\lambda _{1}-k_{c}/2,  \notag \\
Q_{1} &=&2A_{c}L_{1R}\cosh \theta _{1}+\left( \left\vert L_{1}\right\vert
^{2}+A_{c}^{2}\right) \cos \Phi _{1},  \notag \\
Q_{2} &=&2A_{c}L_{1I}\sinh \theta _{1}+\left( \left\vert L_{1}\right\vert
^{2}-A_{c}^{2}\right) \sin \Phi _{1},  \notag \\
\Delta _{1} &=&\left( \left\vert L_{1}\right\vert ^{2}+A_{c}^{2}\right)
\cosh \theta _{1}+2A_{c}L_{1R}\cos \Phi _{1},  \label{para2} \\
x_{0} &=&-\left( \ln \left\vert C_{2}/C_{1}\right\vert \right) /2,\text{ }%
\varphi _{0}=\left[ \arg \left( C_{2}/C_{1}\right) \right] /2,  \notag
\end{eqnarray}%
$\allowbreak $here the subscript $R$ and $I$ represent the real part and
image part, respectively. The parameters $C_{1},C_{2}$ are the arbitrary
complex constants. It should be noted that $A_{c}$, $k_{c}$, $x_{0}$, $\phi
_{0}$, and the complex variable $\lambda _{1}$ are the adjustable parameters
whose values characterize the independent solutions.

When the spin-wave amplitude vanishes, namely $A_{c}=0$, the solution in Eq.
(\ref{onesoliton}) reduces to the solution in the form 
\begin{eqnarray}
m_{x} &=&\frac{\mu _{1}}{\cosh \theta _{1}}\cos \left( \Phi _{1}-\gamma
\right) ,  \notag \\
m_{y} &=&\frac{\mu _{1}}{\cosh \theta _{1}}\sin \left( \Phi _{1}-\gamma
\right) ,  \notag \\
m_{z} &=&\sqrt{1-\frac{\mu _{1}^{2}}{\cosh ^{2}\theta _{1}}},
\label{onesoliton1}
\end{eqnarray}%
where 
\begin{eqnarray*}
\gamma &=&-2\left( 1+\frac{B}{\beta M_{0}}\right) t, \\
\theta _{1} &=&\mu _{1}x+\mu _{1}\nu _{1}t+2x_{0}, \\
\Phi _{1} &=&\nu _{1}x-\frac{1}{2}\left( \mu _{1}^{2}-\nu _{1}^{2}\right)
t-2\varphi _{0}.
\end{eqnarray*}%
The parameters $-2x_{0}/\mu _{1}$ and $2\varphi _{0}/\nu _{1}$ represent the
initial center position and initial phase. The expression (\ref{onesoliton1}%
) describes a magnetization precession characterized by the deviations $\mu
_{1}$ from the ground state, the amplitude $\sqrt{1-\mu _{1}^{2}}$, the
velocity $-\nu _{1}$, and the frequency $\left( \mu _{1}^{2}-\nu
_{1}^{2}\right) /2-2-2B/\left( \beta M_{0}\right) $ which shows that the
magnetic field contribute to precession frequency only. In the other hand,
when $\mu _{1}=0$, the solution (\ref{onesoliton}) reduces to a nonlinear
spin-wave solution%
\begin{eqnarray}
m_{x} &=&A_{c}\cos \varphi ,  \notag \\
m_{y} &=&-A_{c}\sin \varphi ,  \notag \\
m_{z} &=&\sqrt{1-A_{c}^{2}}.  \label{magnon1}
\end{eqnarray}%
Thus the solution (\ref{onesoliton}) describes a soliton solution of
ferromagnet embedded in a nonlinear spin-wave background (\ref{magnon}). The
properties of envelope soliton is characterized by the width $1/\left(
2D_{1R}\right) $, the wave number of soliton $k_{s}=2D_{1I}$, and the
envelope velocity $v_{1}=-\left( D_{1}\delta _{1}\right) _{R}/D_{1R}$. With
the expressions of $D_{1}$ and $\delta _{1}$ we found that the amplitude $%
A_{c}$ and wave number $k_{c}$ of spin wave have the different contribution
to soliton solutions. The width becomes large with the increasing $A_{c}$,
however, becomes decreasing with the increasing $k_{c}$. In the other hand
the velocity of envelope soliton increases continuously with the increasing $%
A_{c}$, however, the increasing at first, then decreasing rapidly with the
increasing $k_{c}$. From Eq. (\ref{para1}) we can directly see that when $%
D_{1I}\delta _{1I}=\delta _{1R}D_{1R}$, the parameters $\theta _{1}$ depends
only on $x$ which implies the envelope velocity $-\left( D_{1}\delta
_{1}\right) _{R}/D_{1R}$ becomes zero, i.e., the magnetic soliton is trapped
in space by the nonlinear spin wave which shows a new technique of
management for the soliton. It should be noted this condition is determined
by the amplitudes $\mu _{1}$, $A_{c}$ of soliton and spin-wave, and the wave
numbers $\nu _{1}$, $k_{c}$ of soliton and spin-wave, respectively. It also
can be seen that the amplitude of soliton in $m_{z}$ is modulated by the
spin wave, and characterized by the spatial and temporal period along the
line $x=-\left( D_{1}\delta _{1}\right) _{R}t/D_{1R}-x_{0}/D_{1R}$, denoted
by $\pi \left( D_{1}\delta _{1}\right) _{R}/\left[ \delta _{1I}\left(
D_{1R}^{2}+D_{1I}^{2}\right) \right] $ and $\pi D_{1R}/\left[ \delta
_{1I}\left( D_{1R}^{2}+D_{1I}^{2}\right) \right] $ obtained from Eqs. (\ref%
{onesoliton}), (\ref{para1}) and (\ref{para2}), respectively. With the
increasing $A_{c}$ the envelope of solution hold the shape of soliton, and
the envelope valley is more deep. Moreover, we can see spin wave move the
center position of soliton.

\section{Two-Soliton interaction modulated by a nonlinear spin wave}

It is well known that soliton has the properties of the physical particles
in the process of collision. Therefore, to investigate the magnetic soliton
collision modulated by a nonlinear spin wave is an very interesting
phenomenon in spin dynamics. To approach this we should obtain the
two-soliton solution of $\mathbf{m}$ from Eqs. (\ref{note}), (\ref{nls2})
and (\ref{Multiso}). Taking the spectral parameter $\lambda =\lambda
_{j}\equiv \mu _{j}/2+i\nu _{j}/2$, $j=1,2$, we obtain the two-soliton
solution of ferromagnet as%
\begin{eqnarray}
m_{x} &=&A_{c}\cos \varphi +\frac{1}{F_{2}}\left( G_{2I}\sin \varphi
+G_{2R}\cos \varphi \right) ,  \notag \\
m_{y} &=&-A_{c}\sin \varphi +\frac{1}{F_{2}}\left( G_{2I}\cos \varphi
-\allowbreak G_{2R}\sin \varphi \right) ,  \notag \\
m_{z} &=&\sqrt{1-\left( A_{c}+\frac{\allowbreak G_{2R}}{F_{2}}\right) ^{2}-%
\frac{G_{2I}^{2}}{F_{2}^{2}}},  \label{twosoliton}
\end{eqnarray}%
where the subscript $R$ and $I$ represent the real part and image part,
respectively, and the other parameters are defined by%
\begin{eqnarray}
\theta _{j} &=&2D_{jR}x+2\left( D_{j}\delta _{j}\right) _{R}t+2x_{0,j}, 
\notag \\
\Phi _{j} &=&2D_{jI}x+2\left( D_{j}\delta _{j}\right) _{I}t-2\varphi _{0,j},
\notag \\
\varphi &=&k_{c}x-\omega _{c}t,  \notag \\
\Delta _{1} &=&\left( \left\vert L_{1}\right\vert ^{2}+A_{c}^{2}\right)
\cosh \theta _{1}+2A_{c}L_{1R}\cos \Phi _{1},  \notag \\
\Delta _{2} &=&\left( \left\vert L_{2}\right\vert ^{2}+A_{c}^{2}\right)
\cosh \theta _{2}+2A_{c}L_{2R}\cos \Phi _{2},  \notag \\
\Delta _{3} &=&\left( \overline{L}_{1}L_{2}+A_{c}^{2}\right) \left[ e^{\frac{%
1}{2}\left( \theta _{1}-i\Phi _{1}+\theta _{2}+i\Phi _{2}\right) }+e^{-\frac{%
1}{2}\left( \theta _{1}-i\Phi _{1}+\theta _{2}+i\Phi _{2}\right) }\right] 
\notag \\
&&+A_{c}\left( \overline{L}_{1}+L_{2}\right) \left[ e^{\frac{1}{2}\left(
\theta _{1}-i\Phi _{1}-\theta _{2}-i\Phi _{2}\right) }+e^{-\frac{1}{2}\left(
\theta _{1}-i\Phi _{1}-\theta _{2}-i\Phi _{2}\right) }\right] ,
\label{para3}
\end{eqnarray}%
\begin{eqnarray}
F_{2} &=&\zeta _{1}\Delta _{1}\Delta _{2}-\mu _{1}\mu _{2}\left\vert \frac{%
\Delta _{3}}{2}\right\vert ^{2},  \notag \\
G_{2} &=&\mu _{2}\zeta _{1}\Delta _{1}f_{2}+\mu _{1}\zeta _{1}\Delta
_{2}f_{1}-\frac{1}{2}\zeta _{2}f_{3}\Delta _{3}-\frac{1}{2}\overline{\zeta }%
_{2}\overline{\Delta }_{3}f_{4},  \notag \\
f_{1} &=&A_{c}\left( L_{1}e^{\theta _{1}}+\overline{L}_{1}e^{-\theta
_{1}}\right) +\left\vert L_{1}\right\vert ^{2}e^{i\Phi
_{1}}+A_{c}^{2}e^{-i\Phi _{1}},  \notag \\
f_{2} &=&A_{c}\left( L_{2}e^{\theta _{2}}+\overline{L}_{2}e^{-\theta
_{2}}\right) +\left\vert L_{2}\right\vert ^{2}e^{i\Phi
_{2}}+A_{c}^{2}e^{-i\Phi _{2}},  \notag \\
f_{3} &=&[A_{c}L_{1}e^{\frac{1}{2}(\theta _{1}+i\Phi _{1})}+A_{c}^{2}e^{-%
\frac{1}{2}(\theta _{1}+i\Phi _{1})}]e^{\frac{1}{2}(\theta _{2}-i\Phi _{2})}
\notag \\
&&+\overline{L}_{2}[L_{1}e^{\frac{1}{2}(\theta _{1}+i\Phi _{1})}+A_{c}e^{-%
\frac{1}{2}(\theta _{1}+i\Phi _{1})}]e^{-\frac{1}{2}(\theta _{2}-i\Phi
_{2})},  \notag \\
f_{4} &=&[A_{c}L_{2}e^{\frac{1}{2}(\theta _{2}+i\Phi _{2})}+A_{c}^{2}e^{-%
\frac{1}{2}(\theta _{2}+i\Phi _{2})}]e^{\frac{1}{2}(\theta _{1}-i\Phi _{1})}
\notag \\
&&+\overline{L}_{1}[L_{2}e^{\frac{1}{2}(\theta _{2}+i\Phi _{2})}+A_{c}e^{-%
\frac{1}{2}(\theta _{2}+i\Phi _{2})}]e^{-\frac{1}{2}(\theta _{1}-i\Phi
_{1})},  \label{para4}
\end{eqnarray}%
\begin{eqnarray}
\zeta _{1} &=&\left\vert \lambda _{2}+\overline{\lambda }_{1}\right\vert
^{2},\text{ }\zeta _{2}=\mu _{1}\mu _{2}\left( \overline{\lambda }%
_{2}+\lambda _{1}\right) ,  \notag \\
D_{j} &=&\sqrt{\left( ik_{c}/2+\lambda _{j}\right) ^{2}-A_{c}^{2}},  \notag
\\
L_{j} &=&-ik_{c}/2-D_{j}-\lambda _{j},  \notag \\
\delta _{j} &=&-i\lambda _{j}-k_{c}/2,  \label{para5} \\
x_{0,j} &=&-\left( \ln \left\vert C_{2,j}/C_{1,j}\right\vert \right) /2,%
\text{ }\varphi _{0,j}=\left[ \arg \left( C_{2,j}/C_{1,j}\right) \right] /2,
\notag
\end{eqnarray}%
where $C_{1,j},C_{2,j}$ are the arbitrary complex constants, $j=1,2$. The
subscript $R$ and $I$ represent the real part and image part, respectively.
It should be noted that $A_{c}$, $k_{c}$, $x_{0,j}$, $\phi _{0,j}$, and the
complex variable $\lambda _{j},$ $j=1,2$, are the adjustable parameters
whose values characterize the independent solutions. In general, the
solution (\ref{twosoliton}) represents the interaction of two one-soliton
solution modulated by the nonlinear spin-wave (\ref{magnon}) of ferromagnet.
The properties of each envelope soliton are characterized by the width $%
1/\left( 2D_{jR}\right) $, the wave number of soliton $2D_{jI}$, and the
envelope velocity $v_{j}=-\left( D_{j}\delta _{j}\right) _{R}/D_{jR}$, $%
j=1,2 $. From Eq. (\ref{para3}) we can directly see that when $D_{jI}\delta
_{jI}=\delta _{jR}D_{jR}$, $j=1,2$, the parameters $\theta _{j}$ depends
only on $x$ which implies the velocity of each envelope becomes zero, i.e.,
the magnetic soliton is trapped in space by the nonlinear spin wave. It also
can be seen that the condition is obviously determined by the amplitudes $%
\mu _{j}$, $A_{c}$ of soliton and spin-wave, and the wave numbers $\nu _{j}$%
, $k_{c}$ of soliton and spin-wave, $j=1,2$, respectively. We also observe
that the amplitude of each envelope soliton in $m_{z}$ is changed by the
spin wave, and characterized by the spatial and time period along the line $%
x=-\left( D_{j}\delta _{j}\right) _{R}t/D_{jR}-x_{0}/D_{jR}$, denoted by $%
\pi \left( D_{j}\delta _{j}\right) _{R}/\left[ \delta _{jI}\left(
D_{jR}^{2}+D_{jI}^{2}\right) \right] $ and $\pi D_{jR}/\left[ \delta
_{jI}\left( D_{jR}^{2}+D_{jI}^{2}\right) \right] $, $j=1,2$, for each
envelope soliton, respectively. With the action of spin wave the envelope of
solution hold the shape of soliton, and the amplitude of each envelope
soliton keeps invariability before and after collision which shows the
elastic collision of two envelope soliton in the nonlinear spin wave
background. Moreover, we can see spin wave move the center position of
soliton oppositely before and after collision.

When $\mu _{j}=0,$ $j=1,2,$ the solution (\ref{twosoliton}) reduces to the
nonlinear wave solution (\ref{magnon}). As the nonlinear spin-wave amplitude
vanishes, $A_{c}=0,$ the two soliton solution (\ref{twosoliton}) reduces to 
\begin{eqnarray}
m_{x} &=&\frac{1}{F_{2}}\left( G_{2I}\sin \gamma +G_{2R}\cos \gamma \right) ,
\notag \\
m_{y} &=&\frac{1}{F_{2}}\left( G_{2I}\cos \gamma -\allowbreak G_{2R}\sin
\gamma \right) ,  \notag \\
m_{z} &=&\sqrt{1-\frac{\allowbreak G_{2R}^{2}+G_{2I}^{2}}{F_{2}^{2}}},
\label{twosoliton1}
\end{eqnarray}%
where%
\begin{eqnarray*}
\gamma &=&-2\left( 1+\frac{B}{\beta M_{0}}\right) t, \\
\theta _{j} &=&\mu _{j}x+\mu _{j}\nu _{j}t+2x_{0}, \\
\Phi _{j} &=&\nu _{j}x-\frac{1}{2}\left( \mu _{j}^{2}-\nu _{j}^{2}\right)
t-2\varphi _{0}, \\
F_{2} &=&\zeta _{1}\cosh \theta _{1}\cosh \theta _{2}-\frac{1}{2}\mu _{1}\mu
_{2}\left[ \cos (\Phi _{1}-\Phi _{2})+\cosh (\theta _{1}\allowbreak +\theta
_{2})\right] , \\
G_{2} &=&\left[ \mu _{2}\zeta _{1}\cosh \theta _{1}-\frac{1}{2}\left( \zeta
_{2}e^{\theta _{1}}+\overline{\zeta }_{2}e^{-\theta _{1}}\right) \right]
e^{i\Phi _{2}} \\
&&+\left[ \mu _{1}\zeta _{1}\cosh \theta _{2}-\frac{1}{2}\left( \zeta
_{2}\allowbreak e^{-\theta _{2}}+\overline{\zeta }_{2}e^{\theta _{2}}\right) %
\right] \allowbreak e^{i\Phi _{1}}.
\end{eqnarray*}%
$\allowbreak $ The solution (\ref{twosoliton1}) describes a general
scattering process of two solitary waves with different center velocities $%
-\nu _{1}$ and $-\nu _{2}$, different phases $\Phi _{1}-\gamma $ and $\Phi
_{2}-\gamma $. Before collision, they move towards each other, one with
velocity $-\nu _{1}$ and shape variation frequency $\Omega _{1}=\left( \mu
_{1}^{2}-\nu _{1}^{2}\right) /2-2-2B/\left( \beta M_{0}\right) $ and the
other with $-\nu _{2}$ and $\Omega _{2}=\left( \mu _{2}^{2}-\nu
_{2}^{2}\right) /2-2-2B/\left( \beta M_{0}\right) $. Asymptotically, the
two-soliton waves (\ref{twosoliton1}) can be written as a combination of two
one-soliton solutions (\ref{onesoliton1}) with different amplitudes and
phases. The asymptotic form of two-soliton solution in limits $t\rightarrow
-\infty $ and $t\rightarrow \infty $ is similar to that of the one-soliton
solution (\ref{onesoliton1}). During collision there is no amplitude
exchange among three components $m_{x}$, $m_{y}$ and $m_{z}$, however, a
phase and the center position change for each magnetization vector soliton.
This interaction between two magnetic solitons is called elastic collision.

\section{Conclusion}

In conclusion, we study the dynamics of the magnetic soliton modulated by a
nonlinear spin-wave in a uniaxial anisotropic ferromagnet. In terms of a
reasonable assumption we transform the Landau-Lifshitz equation into an
equation of the nonlinear type. By means of a straightforward Darboux
transformation one- and two-soliton solutions in nonlinear spin-wave
background are obtained analytically and their properties are discussed in
detail. Our results show that in the background of a nonlinear spin wave the
amplitude of the envelope soliton has the spatial and temporal period. The
soliton can be trapped only in space which is determined by the amplitude
and wave number of the magnetic soliton and the nonlinear spin wave. The
amplitude and wave number of spin wave have the different contribution to
the width, velocity, and the amplitude of soliton solutions. Moreover, we
also observe that the envelope of solution hold the shape of soliton, and
the amplitude of each envelope soliton keeps invariability before and after
collision which shows the elastic collision of two envelope soliton in a
nonlinear spin wave background.

\section{Acknowledgment}

This work was supported the Natural Science Foundation of China under Grant
No. 10647122, the Natural Science Foundation of Hebei Province of China
Grant No. A2007000006, the Foundation of Education Bureau of Hebei Province
of China Grant No. 2006110, and the key subject construction project of
Hebei Provincial University of China.

\end{document}